\documentclass[aps,pra,longbibliography,reprint,showpacs,superscriptaddress,footinbib,citeautoscript]{revtex4-1}

\usepackage{amsmath}
\usepackage{graphicx}
\usepackage{dcolumn}
\usepackage{bm}
\usepackage{amssymb}
\usepackage{epsfig}
\usepackage{color}
\usepackage[normalem]{ulem}
\usepackage[mathlines]{lineno}

\graphicspath{{.}{./figures/}{./EPS/}}

\begin{document}

\title{Superconductivity in the ferromagnetic semiconductor samarium nitride}

\author{E.-M. Anton}
\affiliation{MacDiarmid Institute for Advanced Materials and Nanotechnology, School of Chemical and Physical
Sciences, Victoria University of Wellington, P.O. Box 600, Wellington 6140, New Zealand}

\author{S. Granville}
\affiliation{MacDiarmid Institute for Advanced Materials and Nanotechnology, Robinson Research Institute,
Victoria University of Wellington, P.O. Box 33436, Lower Hutt 5046, New Zealand}

\author{A. Engel}
\affiliation{Physics Institute, University of Z{\"u}rich, Winterthurerstr.\ 190, 8057
Z{\"u}rich, Switzerland}

\author{S.~V. Chong}
\affiliation{MacDiarmid Institute for Advanced Materials and Nanotechnology, Robinson Research Institute,
Victoria University of Wellington, P.O. Box 33436, Lower Hutt 5046, New Zealand}

\author{M. Governale}
\affiliation{MacDiarmid Institute for Advanced Materials and Nanotechnology, School of Chemical and Physical
Sciences, Victoria University of Wellington, P.O. Box 600, Wellington 6140, New Zealand}

\author{U. Z{\"u}licke}
\affiliation{MacDiarmid Institute for Advanced Materials and Nanotechnology, School of Chemical and Physical
Sciences, Victoria University of Wellington, P.O. Box 600, Wellington 6140, New Zealand}

\author{A.~G. Moghaddam}
\affiliation{Department of Physics, Institute for Advanced Studies in Basic
Sciences (IASBS), Zanjan 45137-66731, Iran}

\author{H.~J. Trodahl}
\affiliation{MacDiarmid Institute for Advanced Materials and Nanotechnology, School of Chemical and Physical
Sciences, Victoria University of Wellington, P.O. Box 600, Wellington 6140, New Zealand}

\author{F. Natali}
\affiliation{MacDiarmid Institute for Advanced Materials and Nanotechnology, School of Chemical and Physical
Sciences, Victoria University of Wellington, P.O. Box 600, Wellington 6140, New Zealand}

\author{S. V\'{e}zian}
\affiliation{Centre de Recherche sur l'H\'et\'ero-\'Epitaxie et ses Applications (CRHEA), Centre National de la
Recherche Scientifique, Rue Bernard Gregory, 06560 Valbonne, France}

\author{B.~J. Ruck} \email{ben.ruck@vuw.ac.nz}
\affiliation{MacDiarmid Institute for Advanced Materials and Nanotechnology, School of Chemical and Physical
Sciences, Victoria University of Wellington, P.O. Box 600, Wellington 6140, New Zealand}

\date{\today}

\begin{abstract}

Conventional wisdom expects that making semiconductors ferromagnetic requires doping with magnetic ions,
and that superconductivity cannot coexist with magnetism. However, recent concerted efforts exploring new
classes of materials have established that intrinsic ferromagnetic semiconductors exist and that certain types
of strongly correlated metals can be ferromagnetic and superconducting at the same time. Here we show that
the trifecta of semiconducting behavior, ferromagnetism and superconductivity can be achieved in a single
material. Samarium nitride (SmN) is a well-characterised intrinsic ferromagnetic semiconductor, hosting strongly
spin-ordered 4$f$ electrons below a Curie temperature of $27\,$K. We have now observed that it also hosts a
superconducting phase below $4\,$K when doped to electron concentrations above $10^{21}\,$cm$^{-3}$. The
large exchange splitting of the conduction band in SmN favors equal-spin triplet pairing with $p$-wave symmetry.
An analysis of the robustness of such a superconducting phase against disorder leads to the conclusion that the
4$f$ bands are crucial for superconductivity, making SmN a heavy-fermion-type superconductor.

\end{abstract}

\maketitle

\section{Introduction}

Conventional superconductivity arises due to spin-singlet Cooper pairing of electrons into an
\textit{s}-wave orbital bound state~\cite{Tinkham}. While most elemental and compound superconductors
are of the conventional type, alternative pairing mechanisms have attracted great interest over the
years~\cite{Sigrist1991,Mineev1999,Mackenzie2003}, most recently fuelled by the drive to understand the
origin of high superconducting transition temperatures~\cite{Deutscher}. The fermionic nature of electrons
requires that spin-singlet (spin-triplet) Cooper pairs have even (odd) orbital-bound-state angular momentum.
Hence, unconventional spin-singlet (spin-triplet) superconducting order parameters can have \textit{d}-wave,
\textit{g}-wave, etc.\ (\textit{p}-wave, \textit{f}-wave, etc.) symmetry.

The possibility to form spin-triplet Cooper-pair states underpins the coexistence of magnetism and
superconductivity observed in certain bulk metals~\cite{SaxenaUGe2,Pfleiderer2001,Aoki2001,Pfleiderer2009,
Aoki2011}. The same physics also enables the generation of spin-polarized supercurrents in hybrid structures
of conventional (\textit{s}-wave) superconductors with ferromagnets~\cite{Bergeret2001,Eschrig2003}, thus
opening interesting new opportunities for implementing a spin-based electronics paradigm~\cite{Wolf2001} in
the context of superconductivity~\cite{Linder2015,Eschrig2015,Basaran2015}. The \textit{p}-wave spin-triplet
pairing channel has become prominent as the most likely candidate mechanism at play in real systems.
Furthermore, recent interest has focused on quasi-one-dimensional \textit{p}-wave superconductivity induced
in \emph{semiconductor\/} nanowires with strong-spin orbit coupling through proximity to an \textit{s}-wave
superconductor, because such hybrid systems can be driven into a toplogically nontrivial phase where they
host Majorana excitations~\cite{Lutchyn2010,Oreg2010,Mourik2012,Das2012}. This last development is
only the most recent advance in the long-running drive to further enhance the versatility of semiconductors'
materials properties, e.g., by making them superconducting~\cite{Schaepers,Herrmannsdorfer2009,Blase2009}
or ferromagnetic~\cite{Jungwirth2006,Dietl2010,Dietl2014,Jungwirth2014}, and thus enable the realization and
exploitation of new electronic and magnetic phenomena.

Here we report the presence of superconductivity in the doped intrinsic ferromagnetic semiconductor
SmN, in which magnetism and superconductivity are simultaneously observed in a semiconducting material.
While the coexistence of ferromagnetic order and superconductivity is in itself an interesting phenomenon,
the fact that this coexistence occurs in a semiconductor offers an enhanced level of tunability and potential
for novel device applications. Peculiar to SmN, the large exchange splitting of its conduction and valence
bands~\cite{Larson2007} renders it very likely to harbour fully spin-polarised carriers, thus ruling out any
type of spin-singlet pairing as the origin of the observed superconducting order -- including the putative
inhomogeneous (FFLO) scenario~\cite{Fulde1964,Larkin1965}. Instead, the superconducting state hosted
by SmN is of an unconventional triplet type, most likely exhibiting $p$-wave symmetry. Our work ushers in
an era of new opportunities associated with the ability to control semiconducting, ferromagnetic and
superconducting properties in a single material and, in the process access novel states of quantum
matter~\cite{Editorial2016}.

The paper is organised as follows. In the immediately following Section~\ref{properties}, the basic magnetic
and electronic properties of SmN are introduced. We then provide details about the experimental techniques
employed in our study in Section~\ref{details}. The main  experimental results are presented in Section~\ref{results},
both for thin-film SmN samples and for SmN/GdN superlattices. Section~\ref{theory} is devoted to a theoretical
analysis of the experimental results, and we summarize the main findings of the paper in Section~\ref{summary}.

\section{Review of the magnetic and electronic properties of SmN}
\label{properties}

The past decade has seen significant advances in the growth and passivation of thin films of the rare-earth nitrides
LN, where L denotes a member of the lanthanide series~\cite{Aerts2004,DuanJPC2007,Natali2013b}. In particular,
it is known that almost all of them are semiconductors in both their paramagnetic and ferromagnetic states. In the
same time frame, there has been progress in theoretical treatments of strongly correlated electrons that enabled
reliable band-structure calculations that show good agreement with recent experimental studies~\cite{Preston2007}.
The band structures predicted by the treatments show indirect band gaps, with the N~2$p$ valence band (VB)
maximum at $\Gamma$ and the L~5$d$ conduction band (CB) minimum at $X$. In addition, there are L~4$f$ bands
that cross the valence and conduction bands at energies that vary across the series. GdN, with its exactly half filled
4$f$ shell, has the majority-spin 4$f$ bands about 7~eV below, and the minority-spin bands about 5~eV above the
gap~\cite{DuanJPC2007,Larson2007}. In materials involving lighter L elements, there are empty majority-spin bands
threading the CB, while for LNs with heavier L elements, there are filled minority-spin bands crossing the VB. These
are generally very narrow, heavy-mass bands, though they must hybridise at some level with the 2$p$ and 5$d$ bands.

\begin{figure}[t]
\includegraphics[width=7.5cm]{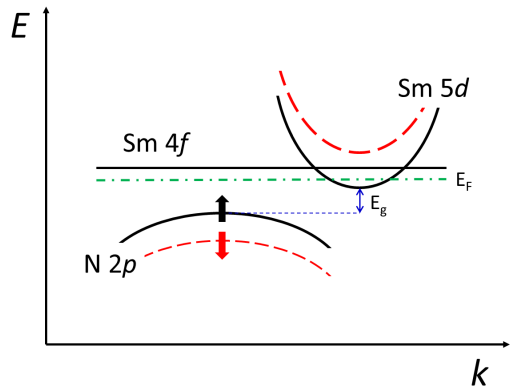}
\caption{\label{fig:bandstructure}
Schematic representation of the band structure of SmN. The valence and conduction
bands are split into majority-spin (black solid curves) and minority-spin (red dashed curves)
by the exchange interaction with filled 4$f$ levels. An empty majority-spin 4$f$ level is shown
as a dispersionless band near the conduction-band minimum. Its exact location, and the size
(or even sign) of the band gap $E_\mathrm{g}$, are not well known. Here the Fermi level
$E_\mathrm{F}$ is shown inside the conduction band so that the SmN is degenerately doped.}
\end{figure}

For GdN, it is now clear that there is a finite band gap~\cite{Trodahl2007,Vidyasagar2013b}, which has facilitated
this material's integration into device structures~\cite{Senapati2011,Kandala2013}. The available experimental data
also support the presence of a finite band gap in SmN~\cite{Natali2013b,Preston2007,AzeemPhD}, although the
data are less complete than for GdN and the possibility of a small band overlap cannot be ruled out at
present~\cite{Morari2015}. In any case, the rare-earth nitrides commonly exhibit high carrier concentrations and
metallic behaviour due to large concentrations of N vacancies, which potentially release three electrons per
defect~\cite{Natali2013b,Lee2015,Punya2011}. The band structure of SmN is represented schematically in
Figure~\ref{fig:bandstructure}, where the metallic behavior is represented by the Fermi level $E_\mathrm{F}$
being located above the CB minimum~\cite{Natali2013b,Leuenberger2005}. This arrangement is consistent
with the experimentally measured carrier concentration of $2\times 10^{21}$~cm$^{-3}$ in the SmN sample
showing superconductivity (see below), which corresponds to a N-vacancy concentration of a few percent.

\begin{figure}[t]
\includegraphics[width=7.5cm]{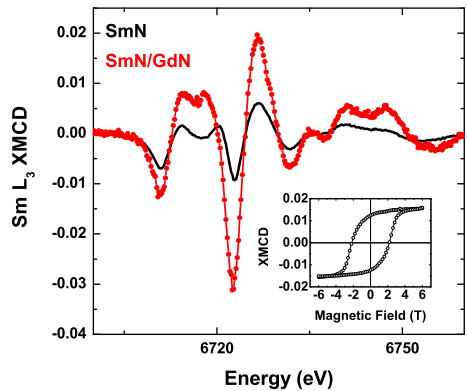}
\caption{\label{fig:XMCD}
X-ray magnetic circular dichroism (XMCD) at the Sm L$_2$ edge in a bulk SmN film and a SmN/GdN superlattice
taken at 15 K and 6T, which measures transitions into conduction-band (CB) states. The large XMCD amplitude
demonstrates the strong exchange splitting between spin-up and spin-down CB states. The XMCD signal is three
times larger in the superlattice sample (cf.\ Refs.~\onlinecite{Anton2013,McNulty2015}), implying an even stronger
CB spin splitting in that case. The inset shows the associated magnetic hysteresis of the bulk SmN film
measured using XMCD, confirming the ferromagnetism.}
\end{figure}

The lanthanide (L$^{3+}$) ions in the nitrides have partially filled 4$f$ shells which harbour the majority of the
magnetic moments. However, unlike transition-metal compounds, the strong spin-orbit coupling and the relatively
weak intra-ion overlap conspire to prevent a full quenching of the orbital magnetic moment. As a result, the net
magnetic moment of these ions can be dominated by either the spin or orbital contribution. SmN holds a special place in the
series~\cite{McNulty2016}. As expected from applying Hund's rules, it has a near zero magnetic moment in the
ferromagnetic phase due to an almost perfect cancellation of the 4$f$ spin magnetic moment by the opposing 4$f$
orbital contribution. Nonetheless its 4$f$ spins are aligned as in any ferromagnet, showing a coercive field of more
than 6~T at 2~K~\cite{Meyer2008,Anton2013}. The evidence for ferromagnetic alignment is unambiguous, including a clear
hysteretic behaviour of the magnetisation, the 4$f$ spin alignment measured by X-ray magnetic circular dichroism
(XMCD) at the $M_{4,5}$ edges, and neutron diffraction studies. The large 4$f$ -- 5$d$ exchange interaction then
drives a large spin splitting of several hundred meV in the Sm 5$d$-dominated conduction band~\cite{Larson2007}.
A direct signature of that spin splitting is provided by XMCD at the Sm $L_{2,3}$ edges. These 2$p$--5$d$ transitions
represent spin-resolved electronic transitions into the conduction band, so that the spectra are to first order
proportional to the size of the spin splitting. The strong XMCD signal shown in Fig.~\ref{fig:XMCD} confirms that the
conduction band is indeed strongly spin-split, and an associated hysteresis further confirms the ferromagnetic ordering
of the Sm moments~\cite{Anton2013}.

\section{Details of experimental methods}\label{details}

\begin{figure}[b]
\includegraphics[width=7.5cm]{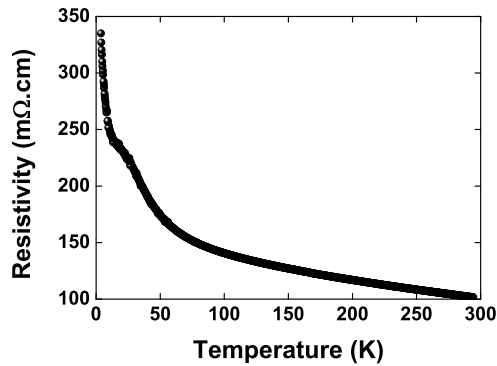}
\caption{\label{fig:SmNstoichRes}
The resistivity of a near-stoichiometric SmN film with low carrier concentration. In this film, the resistivity rises
continuously as the temperature falls, with an anomaly at the Curie temperature signalling the development of
spin splitting at the band edges. There is no sign of an onset of superconductivity in this sample.}
\end{figure}

The films used in this study were prepared by molecular beam epitaxy in our two laboratories, at Victoria University
of Wellington and at CRHEA-CNRS in Valbonne. High purity Sm or Gd metal were evaporated in the presence of
$\sim3\times 10^{-4}$~mbar molecular nitrogen, which react spontaneously with the metal to form the nitride. The
film thicknesses of $\sim100$~nm were determined by Rutherford backscattering spectrometry. The superlattice
comprised layers of 10~nm GdN and 5~nm SmN, determined with a microbalance calibrated by prior Rutherford
backscattering spectrometry. The (111)-oriented epitaxial films were grown at $400-500~^\circ$C on hexagonal
faces of AlN or GaN, which were in turn grown on either commercial c-plane sapphire or (111)~Si. To prevent
oxidation a $\sim50$-nm-thick GaN capping layer was grown on top of the rare-earth nitride films. For the
capping-layer growth, the substrate was held at room temperature and the Ga or Al metal evaporated in the
presence of $3\times10^{-4}$~mbar of activated nitrogen or ammonia. Further details of the growth procedure
can be found elsewhere~\cite{Natali2014,Louarn2009}.

The low energy of formation of nitrogen vacancies dictates that their concentration is of order 1\% at the high
growth temperatures used for the rare-earth nitride layers~\cite{Punya2011}, but reducing the temperature yields
films with lower nitrogen-vacancy doping~\cite{Natali2013b}. Figure~\ref{fig:SmNstoichRes} shows a typical
temperature-dependent resistivity for a SmN sample with low nitrogen-vacancy doping. The data establish
clearly that this near-stoichiometric SmN is a semiconductor, with no sign of superconductivity.

\begin{figure*}[t]
\includegraphics[width=7.5cm]{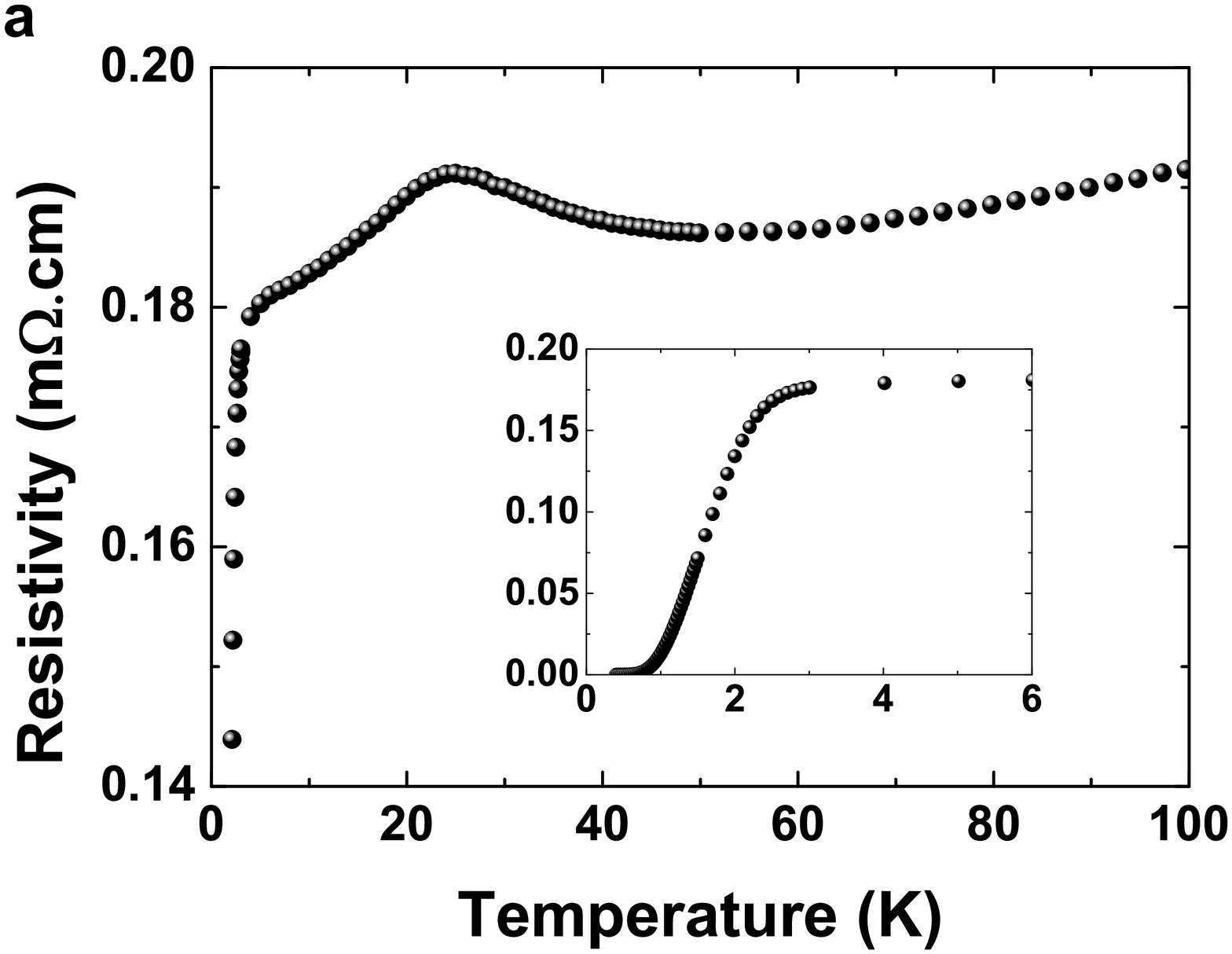}
\includegraphics[width=7.5cm]{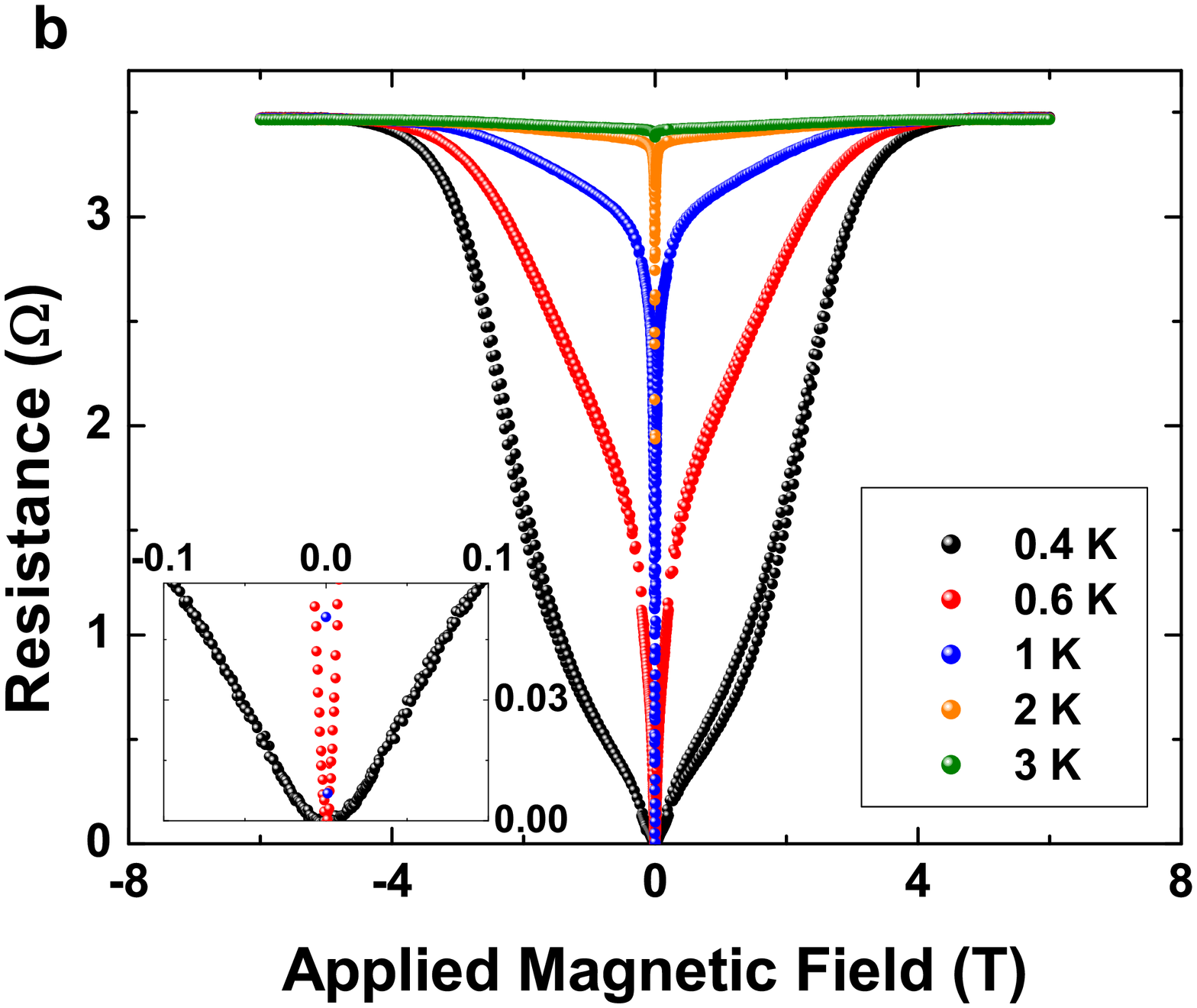}\\
\includegraphics[width=7.5cm]{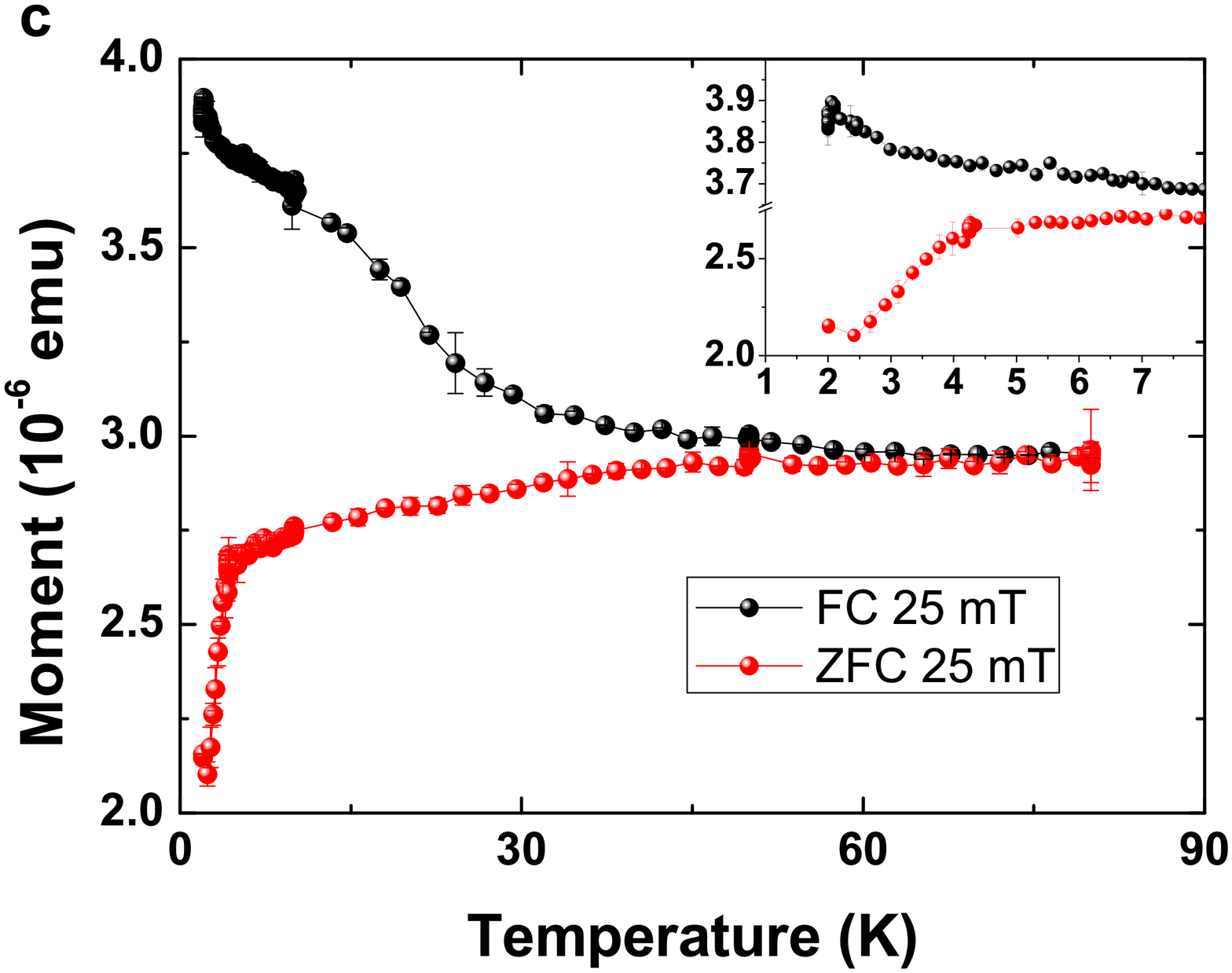}
\includegraphics[width=7.5cm]{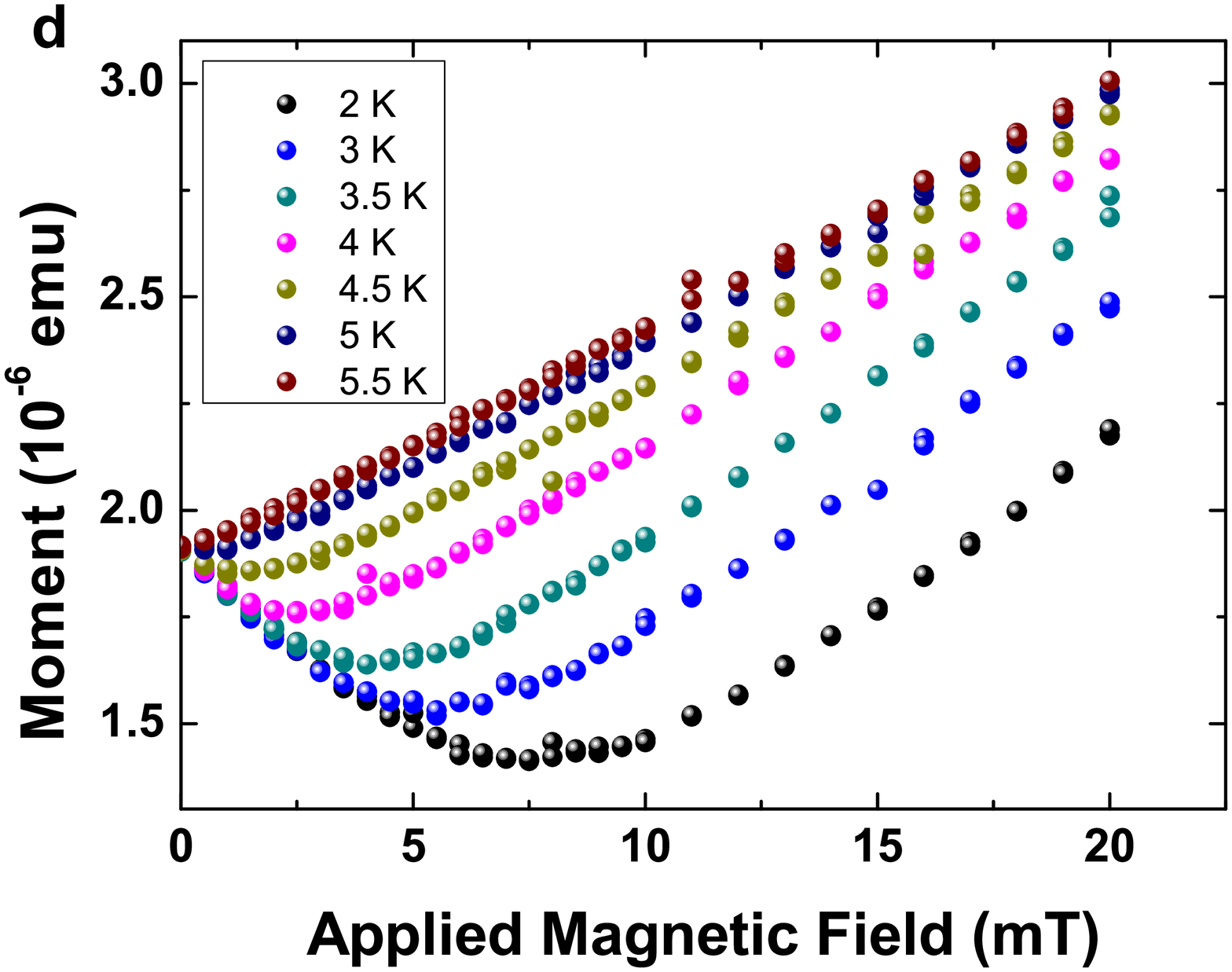}
\caption{\label{fig:SmN}
(a)~Temperature-dependent resistivity of a SmN film, showing an anomaly near
the Curie temperature of 27~K and the onset of superconductivity below about 3~K. The
inset shows the full superconducting transition. (b)~Perpendicular field magnetoresistance
of the sample shown in (a). (c)~Field-cooled (FC) and zero-field-cooled (ZFC) magnetisation data for
a bulk SmN film, showing the ferromagnetic ordering that occurs below 27~K, along with
evidence for the onset of superconductivity below about 4.5~K. (d) Field-dependent
magnetisation showing  diamagnetism at low fields associated with superconductivity.}
\end{figure*}

The resistivity and magnetoresistance data in Figs.~\ref{fig:SmN}(a) and (b) were measured in the van-der-Pauw
geometry in perpendicular field in a Quantum Design PPMS equipped with a He-3 option. For Fig.~\ref{fig:SL}(b), a
PPMS with a horizontal rotator option was used to allow measurements in both parallel and perpendicular fields with
respect to the plane of the film. The resistivity data in Fig.~\ref{fig:SL}(a) and Fig.~\ref{fig:SmNstoichRes} were
measured in a closed-cycle cryostat. The magnetic data were obtained with a Quantum Design MPMS-7 SQUID
magnetometer. To show the diamagnetism in the superconducting state, the measurements were performed in
field-normal configuration because the superconducting penetration depth is larger than the film thickness.

\section{Experimental results}\label{results}

\subsection{Superconductivity in homogeneous SmN}

While investigating electronic and magnetic properties for epitaxial films of the rare-earth-nitride series, we
have found that many of our SmN films, and so far only SmN, display clear superconductivity when sufficiently
heavily doped. Interestingly, the metallic, non-magnetic end-members of the rare-earth nitride series have been
known to be superconducting for some time~\cite{Yamanaka1998,Roberts1976}.

In Fig.~\ref{fig:SmN}(a) we show the temperature-dependent resistance of a homogeneous SmN film that
displays the superconducting transition at low temperature and, at higher temperature, an anomaly at the Curie
temperature. This film, with an electron concentration of  $2\times 10^{21}\,$cm$^{-3}$, exhibits the onset of
superconductivity at $2-3\,$K and reaches zero resistance below $0.7\,$K. The magnetic-field-driven return to
the normal state is complex, as seen in Fig.~\ref{fig:SmN}(b). Strict zero resistance is destroyed in fields as
low as $10\,$mT, but superconductivity still persists in the majority of the film so that the full return to the normal
state does not occur below a field of about $4\,$T. Within the intermediate state, the resistivity displays hysteresis,
seen clearly in the 0.4-K data in Fig.~\ref{fig:SmN}(b), signalling a coexistence of ferromagnetic and superconducting
order parameters, at least in an intermediate mixed state.

\begin{figure*}[t]
\includegraphics[width=7.5cm]{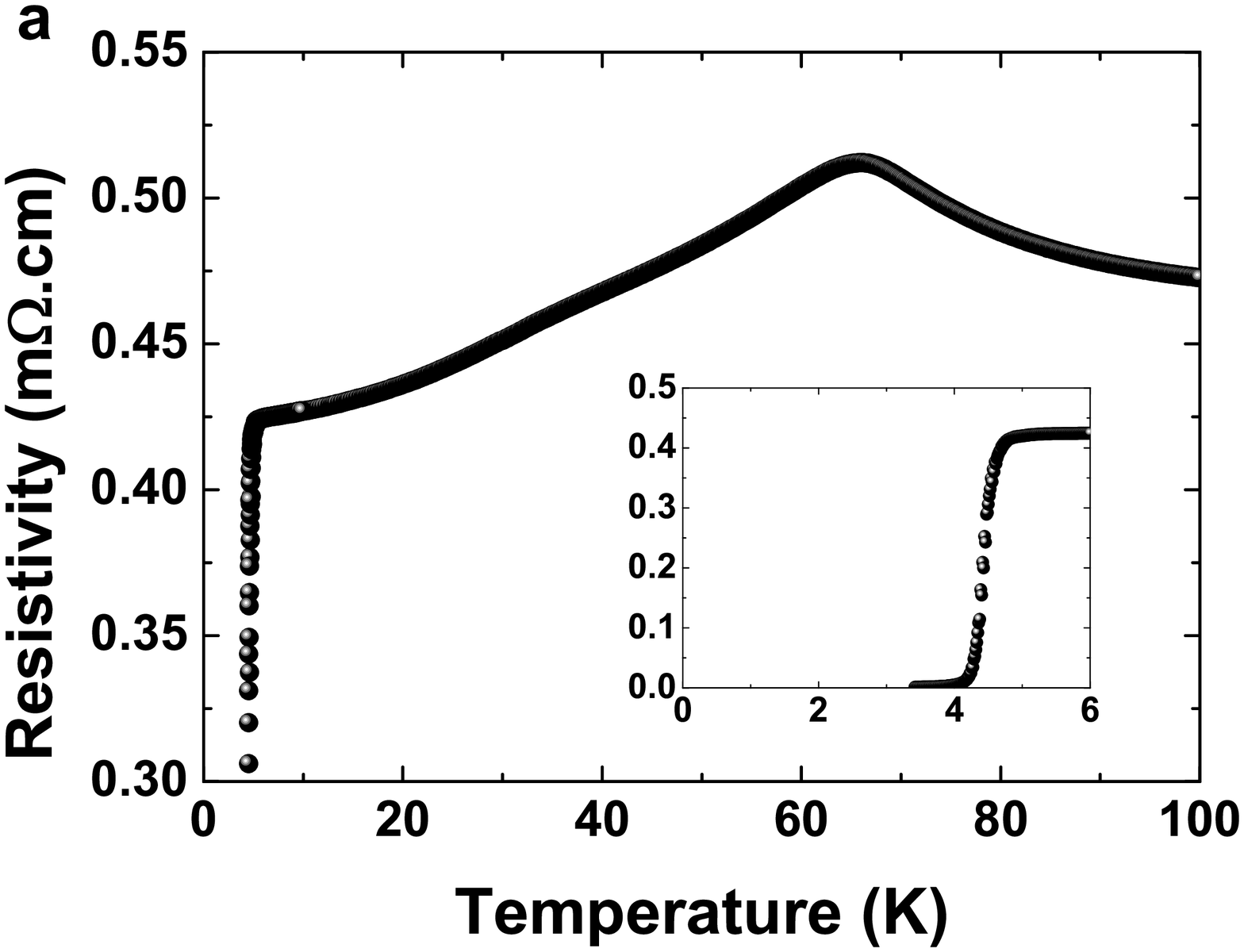}
\includegraphics[width=7.5cm]{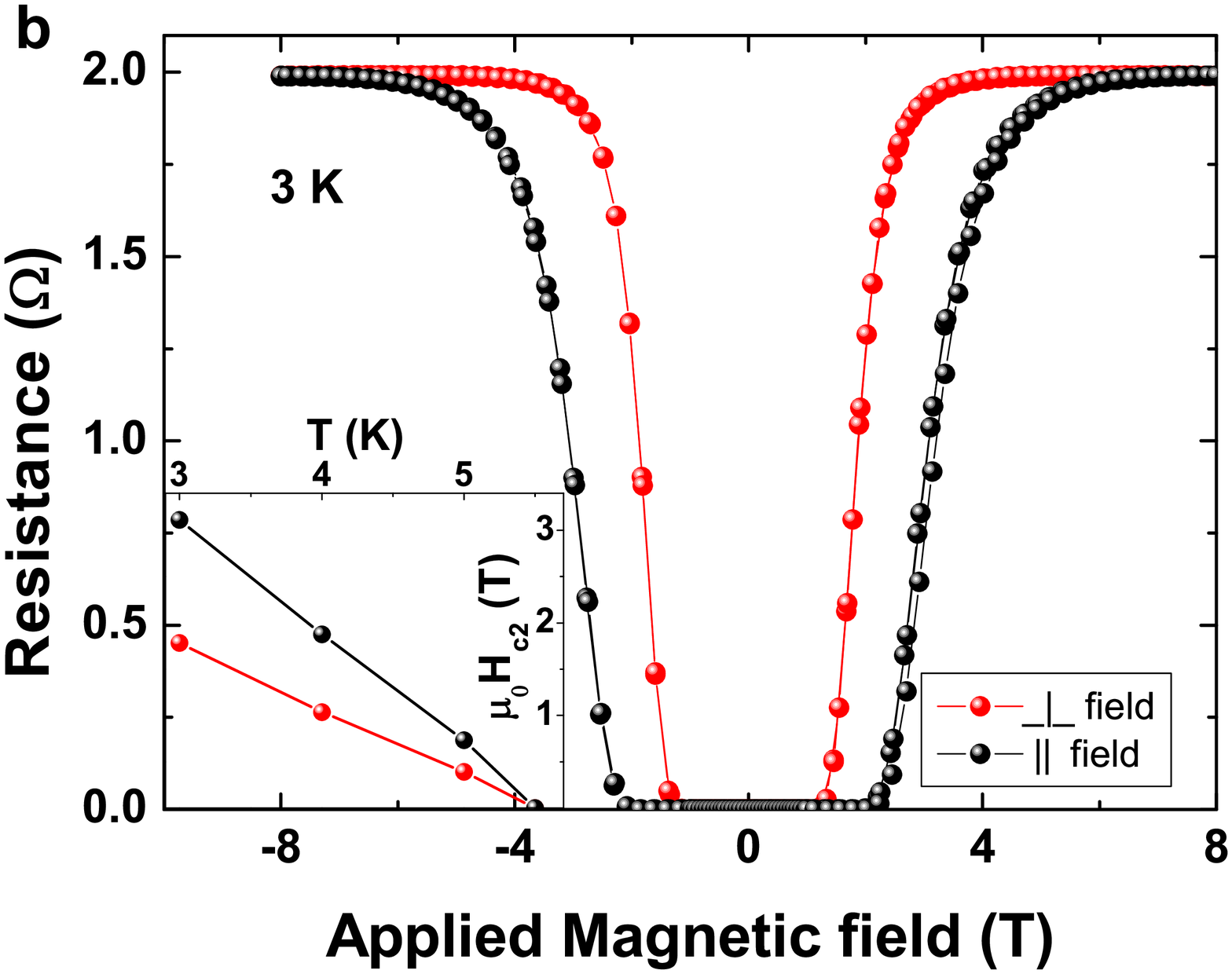}\\
\includegraphics[width=7.5cm]{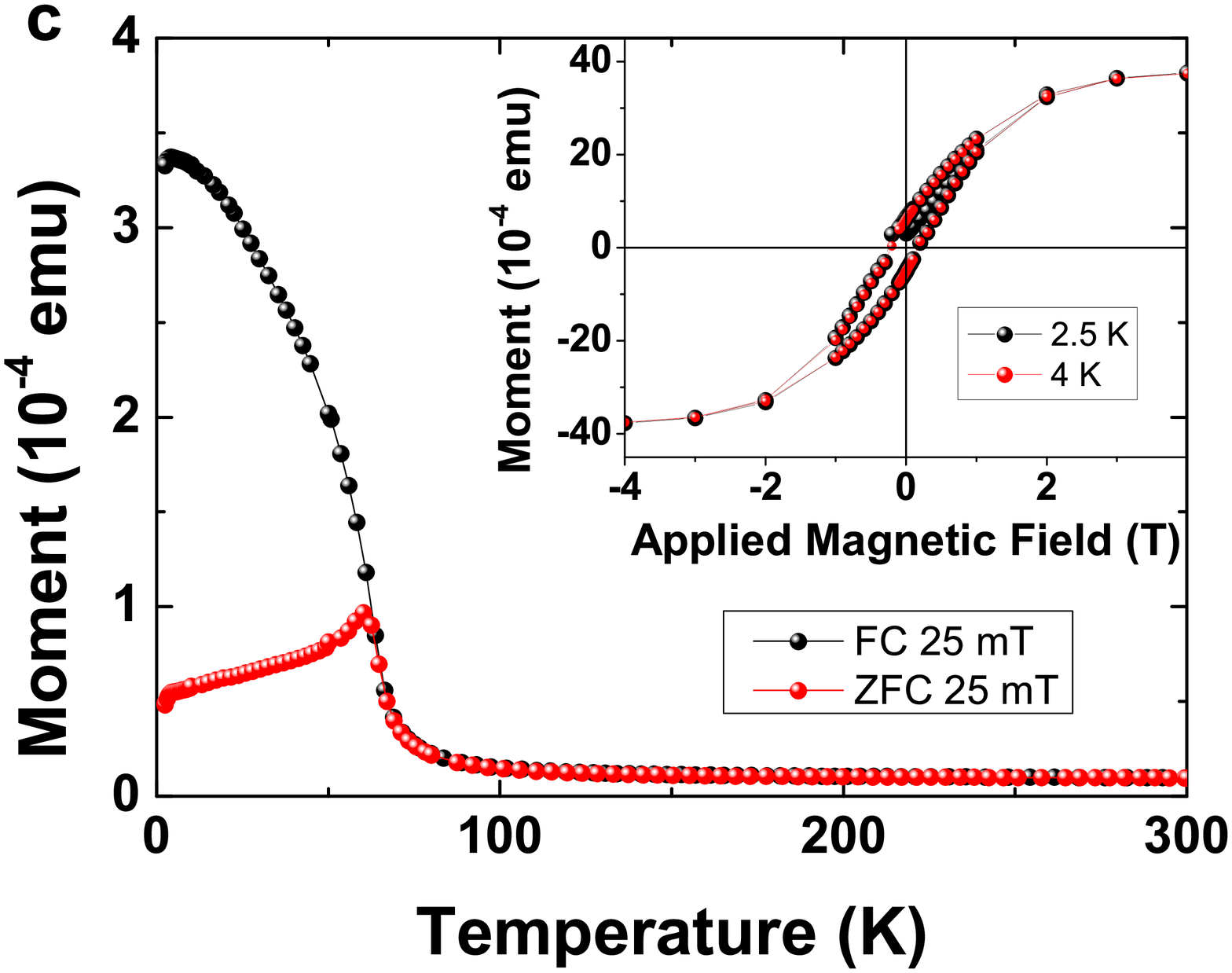}
\includegraphics[width=7.5cm]{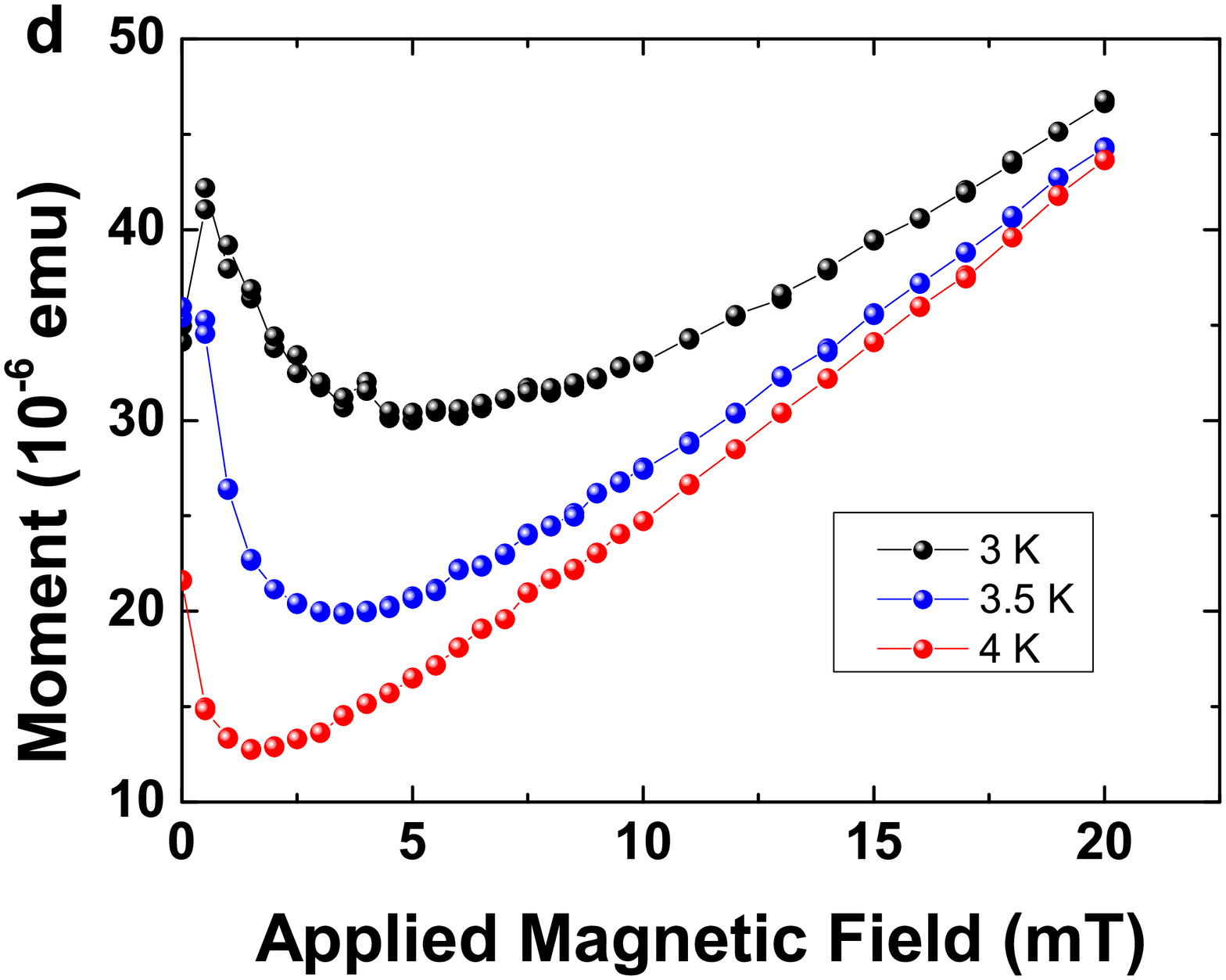}
\caption{\label{fig:SL}
(a)~Temperature dependent resistivity of a SmN/GdN superlattice, showing a much more
abrupt superconducting transition than in bulk SmN films. Anomalies can be seen
corresponding to the Curie temperatures of both the GdN and SmN layers, at $70\,$K and
$30\,$K,  respectively. (b)~Magnetoresistance of the same superlattice in fields applied parallel
(black)  and perpendicular (red) to the layers. The line shape enables straightforward
determination of an upper critical field $\mu_0\,H_\mathrm{c2}$. Inset: The corresponding
upper critical field plotted versus temperature. (c)~The temperature-dependent magnetisation
is overwhelmingly dominated by the GdN. Inset: Hysteresis loops confirming the GdN remains
ferromagnetic even in the temperature range where the SmN is superconducting. (d) The low-field
magnetisation shows a clear diamagnetic response in the superconducting regime that is
superimposed on the large GdN magnetic moment.}
\end{figure*}

In the magnetic measurements shown in Fig.~\ref{fig:SmN}(c), a clear signature of the superconducting
diamagnetic response appears in the zero-field-cooled magnetisation at the lowest temperatures. This
supports the persistence of ferromagnetic order upon entering the superconducting phase, though the
very small magnetisation of SmN limits the sensitivity of magnetisation studies. The coexistence of
ferromagnetic and superconducting order is established even more strongly by the field dependence of
the magnetic moment in Fig.~{\ref{fig:SmN}(d), where the diamagnetic response associated with the
Meissner effect is evident at fields below about $5\,$mT. It is important to note that the diamagnetism is
incomplete, corresponding to only 10\% of the film being in the superconducting state. The incomplete
Meissner effect and the broad superconducting transition suggest that superconductivity nucleates
inhomogeneously such that this bulk film consists of Josephson-coupled superconducting
domains~\cite{Dubi2007,Deutscher}. The inhomogeneity may be associated with grain boundaries and
other structural defects, or it may be of an electronic nature associated with the random distribution of the
nitrogen vacancies that provide the doping. The latter has been observed in GdN films where it leads to
formation of magnetic polarons~\cite{Natali2013a}.

\subsection{Superconductivity in a SmN/GdN superlattice}

In order to investigate further the coexistence of superconductivity and spin-split bands, we have studied
a superlattice consisting of SmN layers in direct contact with strongly ferromagnetic GdN in a
12$\times$(10-nm GdN/5-nm SmN) superlattice. In this context, it is important to recognize that GdN has
never shown any propensity for superconductivity. However, GdN has one of the strongest spin alignments
known, with a ferromagnetically aligned $7\,\mu_\mathrm{B}$ moment on each Gd$^{3+}$ ion below $70\,$K
that resides purely in its spin as the total orbital angular momentum vanishes for this half-filled 4$f$
shell~\cite{Natali2013b}. X-ray magnetic circular dichroism (Fig.~\ref{fig:XMCD}) shows that the ferromagnetic
exchange across the SmN/GdN interface increases the XMCD signal by a factor of 3, i.e., the spin splitting in
the conduction band of the thin SmN layers is strongly enhanced by the proximity to the ferromagnetic GdN
layers~\cite{Anton2013,McNulty2015}.

Transport measurements shown in Figs.~\ref{fig:SL}(a) and (b) reveal a superconducting phase that develops
even more strongly in the superlattice, despite the enhanced spin splitting. Interestingly, not only is the critical
temperature higher in these layers, but also the critical field is enormously enhanced (see Fig.~\ref{fig:SL}(b)),
and zero resistance is maintained now to fields as large as $2\,$T. The coherence length implied by that critical
field is $\sim 10\,$nm~\cite{Tinkham}, i.e., much smaller than the thickness of the homogeneous film above but
comparable to the SmN-layer thickness in the superlattice. The temperature-dependent magnetisation of the
superlattice, shown in Fig.~{\ref{fig:SL}}(c), is dominated by the very strong ferromagnetic moment of GdN,
which has a Curie temperature of $70\,$K. Similar to Fig.~{\ref{fig:SmN}}(c), Fig.~{\ref{fig:SL}}(c) also shows
a diamagnetic anomaly at the lowest temperatures, confirming the existence of the superconducting phase.
Figure~{\ref{fig:SL}(d) displays that diamagnetic response at low fields, with a low-field slope that, in the
superlattice case, corresponds to full diamagnetism in the SmN layers. Hence, again the superconducting
signature is exhibited much more strongly in the superlattice, despite its more strongly spin-split conduction
band.

\section{Theoretical analysis: Spin-triplet heavy-fermion superconductivity}\label{theory}

The experimental data clearly indicate a coexistence of superconductivity and ferromagnetism. The large
exchange splitting of the conduction and valence bands in SmN renders it a half-metallic ferromagnetic
semiconductor at the doping levels realized in our samples. As a result, only majority-spin electrons are
available to form Cooper pairs, and the order parameter will be a fully polarized spin-triplet state
$|\!\!\!\uparrow\uparrow\rangle$. Due to the fermionic nature of electrons, the orbital part of the order
parameter needs to possess odd parity in space, which implies an odd orbital-angular-momentum quantum
number $l=1,3,\dots$ or, using the commonly adopted atomic convention, \textit{p}, \textit{f}, $\dots$
symmetry. Most probably, the lowest value of Cooper-pair angular momentum compatible with the required
order-parameter symmetry will be realised, and therefore we assume a \textit{p}-wave pairing. The associated
pair potential $\Delta_{\uparrow\uparrow}(\mathbf{k})$ can, in principle, have the following possible states:
\begin{align}
\Delta_{\uparrow\uparrow}(\mathbf{k})&= \Delta_0 \times \left\{ \begin{array}{cl}
\frac{k_z}{k_\mathrm{F}} & \mbox{$p_z$-like symmetry} \\[0.3cm]
\frac{k_x\pm i k_y}{k_\mathrm{F}} & \mbox{$p_x \pm i p_y$-like symmetry}
\end{array} \right. .
\end{align}
For both these possibilities, the magnitude of the order parameter is given by $\Delta_0\approx 2 \hbar
\Omega_\mathrm{c} \exp\left(-D/\lambda \right)$, where $D$ is the number of spatial dimensions, $\hbar
\Omega_C$ the cut-off energy for the attractive interaction, and $\lambda$ the dimensionless coupling
constant for the \textit{p}-wave pairing channel. The experimental data do not give us enough information
to infer the origin of the pairing interaction. However, the large splitting between majority and minority
bands allows us to conclude that spin fluctuations play no role. This leaves the combination of
electron-phonon and electron-electron interactions as the most likely cause for Cooper pairing in SmN.
The sharper superconducting transition in the superlattice indicates a cleaner sample and/or
interface-enhanced superconductivity. On the other hand, the broad superconducting transition for the
thin film can be associated with the formation of domains, which is consistent with the measured incomplete
diamagnetic response. See Fig.~\ref{fig:SmN}(d).

An important issue for \textit{p}-wave superconductivity is its weakness with respect to disorder. It has
been established that, for all possible forms of the order parameter in any spatial dimension, the critical
temperature $T_\mathrm{c}$ is suppressed with respect to the one in the absence of disorder
$T_{\mathrm{c0}}$  according to the universal formula~\cite{Mineev1999}
\begin{align}
\label{eq:general}
\ln\frac{T_\mathrm{c0}}{T_\mathrm{c}}=\psi\left(\frac{1}{2}+ \frac{\hbar}{4 \pi \tau
k_\mathrm{B} T_\mathrm{c}}\right) -\psi \left(\frac{1}{2}\right) \quad,
\end{align}
where $\psi$ denotes the Digamma function, $\tau$ the quasiparticle scattering time and $k_\mathrm{B}$
the Boltzmann constant. From the experimentally measured resistivity $\rho=0.18\,$m$\Omega$cm and
carrier density $n= 2\times 10^{21}\,$cm$^{-3}$ for the thin film, we can extract the scattering time $\tau$
as a function of the effective mass $m$ by means of the Drude formula, obtaining $\tau= m / (n q^2 \rho)$,
where $q$ is the electron charge. Using $\tau$ and the experimentally measured $T_\mathrm{c} \approx 3
\,$K we can infer what the critical temperature would be in the absence of disorder. If we use for the effective
mass a value typical for a 5$d$-band $m_{5d}\lesssim 0.5\, m_0$, with $m_0$ being the electron mass in
vacuum,  we obtain a highly unrealistic value for $T_\mathrm{c0}\approx 800\,$K. In order to obtain a
value in the realistic range $T_{\mathrm{c0}} < 30\,$K, the effective charge-carrier mass $m$ needs to be
larger than $15\,m_0$, see Fig.~\ref{fig:Tc0EffeMass}. Therefore, we must therefore conclude that the
4$f$ band is crucial for superconductivity in SmN, making it a heavy-fermion superconductor. We note at
this point that such a large value for the effective mass for the quasi-electrons may necessitate the consideration
of non-adiabatic corrections to the electron-phonon coupling~\cite{Pietronero1995}.

\begin{figure}[t]
\includegraphics[width=7.5cm]{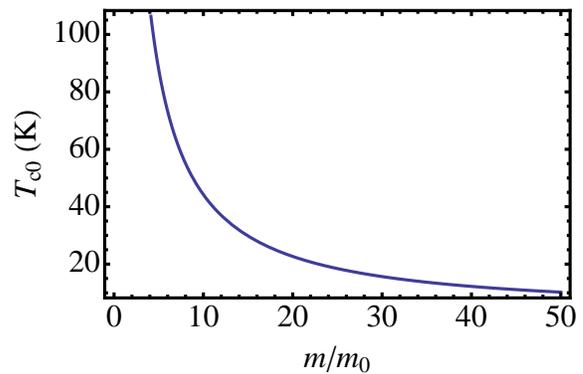}
\caption{\label{fig:Tc0EffeMass}
Dependence of the critical temperature $T_\mathrm{c0}$ in the clean system on the
superconducting charge-carriers' effective mass, $m$, as inferred from the measured
$T_\mathrm{c}\approx 3\,$K, the normal-state resistivity and the carrier density  according
to the universal theory for pair breaking in a \textit{p}-wave superconductor. Here $m_0$
denotes the electron mass in vacuum. A highly unrealistic value for $T_\mathrm{c0}\gtrsim
800\,$K would be associated with the 5$d$-band effective mass $m_{5d}\lesssim 0.5\, m_0$.
Hence we conclude that carriers from the very weakly dispersing 4$f$ band of SmN, or possibly
its hybridisation with the 5$d$ band, must form the superconducting condensate.}
\end{figure}

As a consistency check,  we  assume phonon-mediated pairing and, taking a realistic value of the  Debye
temperature for SmN~\cite{Granville2009} $T_D=\hbar \Omega_c/ k_{\textrm{B}}=600\,$K and an effective
mass of $m= 15\, m_0$, we find $\lambda\approx 0.8$. Such a value corresponds to a physically realistic
strong coupling situation.

\section{Summary and outlook}\label{summary}

In summary, we consistently find superconductivity below a temperature of $5\,$K in heavily doped
ferromagnetic SmN, but not in any of the other rare-earth-nitride films that we have grown. As previous
work has established the unique property of SmN to have a nearly vanishing net magnetic moment
coexisting with a large spin splitting in the conduction band, it can be inferred that the superconductivity
resides in a majority-spin band, implying triplet pairing. SmN is further unusual in the predicted location
of the very narrow majority-spin $4f$ band at the bottom of the conduction band. The full set of
experimental results supports a scenario of SmN being a heavy-fermion spin-polarised superconductor
with \textit{p}-wave triplet pairing.

The discovery of a superconducting ferromagnetic semiconductor paves the way to explore an entirely
new technological paradigm of semiconductor super-spintronics. The ability to adjust the density and
type of charge carriers in this material, the wide range of possibilities for integrating it into heterostructures
with conventional semiconductors or other members of the rare-earth-nitride series, some of which are
expected to exhibit topological order~\cite{Garrity2014,Li2015}, and the unconventional properties of
\textit{p}-wave superconducting phases creates a versatile platform for engineering new quantum
phases of matter with potential for revolutionizing microchip design.

\begin{acknowledgments}

Funding for this work was provided by the Marsden Fund (contract no.\ VUW1309) and the MacDiarmid
Institute for Advanced Materials and Nanotechnology, which is a New Zealand Centre of Research
Excellence. We acknowledge support from GANEX (ANR-11-LABX-0014). GANEX belongs to the
publicly funded Investissements d'Avenir program managed by the French ANR agency. E.-M.A.\ was
supported by a Feodor Lynen Research Fellowship from the Alexander von Humboldt Foundation. A.G.M.\
thanks the School of Chemical and Physical Sciences at Victoria University of Wellington for hospitality. We
thank L.~Figueras for providing the sample and resistivity data on the near-stoichiometric SmN, J.~Kennedy
and P.P.~Murmu for Rutherford backscattering spectrometry measurements, and A.~Schilling, S.~Wimbush
and J.~Storey for discussions.

\end{acknowledgments}

%

\end{document}